\documentclass[twocolumn,showpacs,amsmath,amssymb]{revtex4}
\usepackage{amssymb}

\usepackage{latexsym}
\usepackage{graphicx}
\usepackage{slashed}
\usepackage{pifont}

\newcommand{\req}[1]{Eq.\,(\ref{#1})}

\newcommand{\beqn}{\begin{equation}}
\newcommand{\eeqn}{\end{equation}}

\newcommand{\tr}{\mathrm{tr}\,}

\newcommand{\Imag}{\mathtt{Im}\,}
\newcommand{\Veff}{\ensuremath{V_{\mathrm{eff}}}}

\newcommand{\Fumn}{F^{\mu\nu}}

\newcommand{\magE}{|\vec E|}
\newcommand{\calE}{{\cal E}}
\newcommand{\TEH}{T_{\rm M}}
\newcommand{\THU}{T_{\rm U}}
\newcommand{\fracg}{\frac{g}{2}}

\begin{document}
\title{Acceleration and vacuum temperature}
\author{Lance Labun and Johann Rafelski}
\affiliation{Department of Physics, University of Arizona, Tucson, Arizona 85721, USA}

\date{July 21, 2012} 

\begin{abstract} 
The quantum fluctuations of an ``accelerated'' vacuum state, that is
vacuum fluctuations in the presence of a constant electromagnetic field, can
be described  by the temperature $\TEH$.  Considering $\TEH$ for the 
gyromagnetic factor $g=1$ we show that $\TEH(g=1)=\THU$, where
$\THU$ is the Unruh temperature experienced by an accelerated observer.
We conjecture that both particle production and nonlinear field 
effects inherent in the Unruh accelerated observer case are described 
by the case $g=1$ QED of strong fields.
We present rates of particle production for $g=0,1,2$ and show that the
case $g=1$ is experimentally distinguishable from $g=0,2$. Therefore, 
either  accelerated observers are distinguishable from accelerated 
vacuum or there is unexpected  modification of the theoretical framework.
\end{abstract}

\maketitle
\section{Introduction}
A detector in a matter- and field-free spacetime undergoing constant acceleration $a_{\rm U}$ is found to be embedded in a thermal background at the Unruh temperature ($\hbar =c=1=k_{\rm B}$)
\beqn\label{TUH}
\THU=\frac{a_{\rm U}}{2\pi}.
\eeqn
The statistics of the thermal distribution are bosonic considering the vacuum of a scalar particle~\cite{Unruh:1976db,Crispino:2007eb} and fermionic in the vacuum of a Fermi particle~\cite{Candelas:1978gg}. In other words, the free and unstructured vacuum fluctuations appear to an accelerated observer as having an effective temperature $\THU$ with statistics corresponding to the fluctuation of either Fermi or Bose type.

A complementary effect was recognized by M\"uller {\it et al}.~\cite{Muller:1977mm} who found that the structured vacuum fluctuations induced by an exactly constant electric field $\calE$ (or magnetic field) can be understood as a thermal background  characterized by the temperature parameter
\beqn\label{TEH}
\TEH=\frac{e\calE}{m\pi}\,.
\eeqn
$\TEH$ arises from the exact solution introduced by Heisenberg and Euler~\cite{Heisenberg:1935qt}  and generalized by Schwinger~\cite{Schwinger:1951nm} of vacuum fluctuation properties for constant electromagnetic fields in QED evaluated at lowest order in $\alpha$.

Since an electric field accelerates all charged particles and, in particular, the electron-positron pairs whose fluctuations are considered, it is natural to introduce the global acceleration  $a_v=e\calE/m$~\cite{Greiner:1985ce} (see p.569 ff) and consider this equivalent to an ``accelerated quantum vacuum'' state.   A succinct discussion   is found in the work of Pauchy Hwang and Kim~\cite{PauchyHwang:2009rz}.

Writing $T_M$ in terms of $a_v$ shows a proportionality different by a factor two from the Unruh temperature,
\beqn\label{TEHa}
\TEH=\frac{e\calE}{m\pi} =\frac{a_v}{\pi}=2\THU\,.
\eeqn
The factor two in the temperature is not the only difference between the accelerated vacuum and accelerated observer.  For the case of the   accelerated  vacuum, M\"uller {\it et al}.~\cite{Muller:1977mm} show the associated thermal distribution to be  opposite expectation, being bosonic for spin-1/2 electron fluctuations and fermionic for spin-0 charged particle fluctuations.

The difference between the physical conditions giving rise to the Unruh and M\"uller temperatures is whether it is the observer or the vacuum state that is accelerated. While frame independence of physics phenomena is assured for inertial observers, there is no imperative need for the two cases we consider, accelerated observer and accelerated vacuum, to yield equivalent results.  The two different acceleration cases can be treated by similar methods~\cite{Brout:1995rd}, yet there is difference in outcomes by a factor two highlighted in~\req{TEHa}.  Appearance of two different results suggests new physics content regarding the description of acceleration in terms  of the two reference views, accelerated observer or accelerated vacuum, and suggests that these views are not equivalent, no matter how small the acceleration is. Our objective is to improve the understanding about the origin of this discrepancy and to show that in a special new case this discrepancy disappears.

In QED, the structure of vacuum fluctuations is encoded in the effective action, from which one derives spontaneous particle creation and the associated temperature.  The difference arises in connection with the spin and statistics of the particle. Therefore, we study the structure of the QED vacuum fluctuations in the presence of strong fields for different values of the $g$ factor. We show that the specific value $g=1$ reconciles the temperatures and statistics and discuss pair production in strong fields which can help distinguish the accelerated observer from accelerated vacuum state.
\\[-0.2cm]

\section{Temperature of Electron Fluctuations}
Separate conservation of charge-convective and spin currents means that for any particle the value of the gyromagnetic ratio  $g$ can be arbitrary. For pointlike electrically charged leptons, quantum corrections result in $g-2=\alpha/\pi+...$, and composite spin-1/2 particles have values which can significantly differ from the Dirac value $g=2$.

The dynamics of a particle $\psi$ with arbitrary $g$ is generated by the equation of motion
\beqn\label{EoMg}
\left[D^2+m^2-\frac{g}{2}\frac{e\sigma_{\mu\nu} \Fumn}{2}\right]\psi=0,
\eeqn
where $D=\partial+ieA$ is the covariant derivative, $\Fumn$ the electromagnetic field strength tensor and $\sigma_{\mu\nu}=(i/2)[\gamma_{\mu},\gamma_{\nu}]$. Equation~\eqref{EoMg} comprises a doubling of dynamical components since the ``squared'' equation commutes with $\gamma_5$. For the specific case $g=2$, one can cast \req{EoMg} in  the form of the product of two Dirac equations with $\pm m$.  We   will explicitly show the number of physical degrees of freedom.  The effect of $g$ on the vacuum fluctuations is determined computing the effective potential
\beqn\label{Veffdefn}
\Veff=-\frac{i}{2}\tr\ln\left[D^2+m^2-\frac{g}{2}\frac{e\sigma_{\mu\nu} \Fumn}{2}\right].
\eeqn

The Schwinger proper time method~\cite{Schwinger:1951nm} can be applied to evaluate \req{Veffdefn}   and one finds for $|g|\le 2$
\beqn\label{Veffab}
\Veff = \frac{\gamma_s}{32\pi^2}\!\!\int_{0}^{\infty}\!\!\!\!\!e^{-im^2u}\!\left(\frac{a u\cosh(\frac{g}{2}a u)}{\sinh(au)}\frac{b u\cos(\frac{g}{2}b u)}{\sin (bu)}-1\!\!\right)\!\frac{du}{u^{3}}
\eeqn
in which $\gamma_s$ counts the number of degrees of freedom.  With only bosonic particle and antiparticle degrees of freedom  $\gamma_s=-2$ for $g=0$.  When $g=2$, we have spin-1/2 Dirac fermions, and counting spin degrees of freedom, $\gamma_s=+4$.  The $-1$ inside the parentheses removes the field-independent constant.  In \req{Veffab}, we use $a$ the electriclike and $b$ the magneticlike eigenvalues of $e\Fumn$, which are related to the field strengths by
\begin{align}\label{invar}
a^2-b^2=e^2(\vec E^2-\vec B^2)~~~{\rm and}~~~(ab)^2=e^4(\vec E\cdot\vec B)^2.
\end{align}
The $a$ eigenvalue is electriclike because $a\to e\magE$ in the limit $b\to0$, and similarly $b\to e|\vec B|$ in the limit $a\to 0$. 
 
We discuss here the temperature and statistics for the case of an electric-only field; a transformation similar to that detailed below is possible for the general case \req{Veffab}~\cite{Labun:2008qq}.  For an electric-only field of strength $\calE\equiv\magE$, the $b\to 0$ limit of \req{Veffab} yields
 \beqn\label{Veff}
\Veff = \frac{\gamma_s}{32\pi^2}\!\!\int_{0}^{\infty}\!e^{-im^2u}\left(\frac{e\calE u\cosh(\frac{g}{2}e\calE u)}{\sinh e\calE u}-1\right)\frac{du}{u^{3}}.
\eeqn
Transforming $\Veff$ to a statistical format proceeds via meromorphic expansion of the integrand of \req{Veff}~\cite{Muller:1977mm}.  We introduce the identity
\begin{align}\label{meroexp1} 
1-\frac{z\,\cosh(zy)}{\sinh (z)} 
=&-2z^2\!\sum_{n=1}\frac{\cos n\pi(y+1)}{(n\pi)^2} \\ \notag
&\!+2z^4\!\sum_{n=1}\!\frac{\cos n\pi (y+1)}{(n\pi)^2(z^2+(n\pi)^2)},\ |y|\le 1.
\end{align}
The first term ($\propto z^2$) is identified as the logarithmically divergent contribution and displays the renormalization of charge.

The finite (regularized and renormalized) effective potential is obtained by inserting only the second term of \req{meroexp1} in the integrand of \req{Veff}.
Transforming for $|g|\le 2$ the variable $u\to -inu\pi /e\calE=-inu/m\TEH$, 
\beqn\label{Veff-step1}
\Veff \!\! = \!\!
\frac{\gamma_sm^2\TEH^2}{32\pi^2}\!\!\int_0^{\infty}\!\!\!\frac{2u\:du}{u^2\!-\!1\!+\!i\epsilon}\sum_{n=1}^{\infty}
\frac{e^{-nu\frac{m}{\TEH}}}{n^2}\cos\left(n\pi(\fracg+1)\right)
\eeqn
Note that we have   rotated the integration contour onto the real axis and  defined the integration contour in accordance with the assignment 
\beqn\label{m2minus}
m^2\to m^2-i\epsilon\equiv m^2_-,
\eeqn
which defines the imaginary part discussed further below. While the real part of $\Veff$ controls nonlinear electromagnetic field-field interactions,   its imaginary part controls the rate at which the electromagnetic field decays into electron-positron pairs.

Setting $g=2$ for a spin-1/2 (Dirac) electron, $\cos 2n\pi = 1$ for all $n$, and setting $g=0$ for a spin-0 electron, $\cos n\pi = (-1)^n$ producing an alternating sum.  In each case, integrating by parts twice and summing the series yields the results of M\"uller {\it et al}.~\cite{Muller:1977mm} which for $|g|\le 2$ we present as
\beqn\label{Veffggen}
\Veff = \frac{\gamma_sm^2\TEH}{64\pi^2}\!\!\int_0^{\infty}\!\!\!\!dE \ln(E^2\!-\!m^2_-)\!\sum_{\pm}\ln(1+e^{\pm i\pi\fracg}e^{-E/\TEH}).
\eeqn
The sum over $\pm$ ensures the distribution is real so that the imaginary part arises only from the branch cut in the first log factor. The exponential weights of the terms in the series in \req{Veff-step1} generate for integer values of $g$ an exact  thermal distribution, and the statistics of the distribution are determined by the phase of the terms in the series.

For $g=1$ (and, more generally, for any odd integer value of $g$) summing in \req{Veffggen} over $\pm$ simplifies  to
\beqn\label{Veffg1}
\Veff\Big|_{g=1} \!\!\!\!= \frac{\gamma_sm^2\THU}{32\pi^2}\!\!\int_0^{\infty}\!\!\!\!dE \ln(E^2\!-\!m^2_-)\ln(1+e^{-E/\THU})
\eeqn
exhibiting in the second log factor a thermal fermionic distribution controlled by the Unruh temperature, $\THU$.  The effective potential of a ``classical'' spinning electron with $g=1$ in a constant field thus has the format of a thermodynamic potential with the temperature parameter and statistics in agreement with expectations based on the result obtained for an accelerated observer in the (unaccelerated) vacuum of a fermion field.

We thus find that when the gyromagnetic moment of the electron is that of the ``classical'' spinning particle $g=1$, the differences disappear between an accelerated observer and an accelerated vacuum in both temperature and statistics. This situation is summarized in Table~\ref{tab:accelT}.  It seems that reconciliation of the physics arising under Unruh and M\"uller experimental conditions implies  that we can   no longer  distinguish an accelerated observer from an accelerated vacuum state. However, in our opinion, one must take the evaluation for $g=1$ as a new method to compute the known result attributed to the accelerated observer case.\\[-0.2cm]

\begin{table}
\begin{tabular}{c|c|c||c|c|c}
& \multicolumn{2}{c||}{Detector acceleration} & \multicolumn{3}{|c}{Constant Electric Field}\\
& \multicolumn{2}{c||}{ $a_{\rm U}$ relative to  vacuum} & \multicolumn{3}{|c}{acceleration $a_v=e\calE/m$} 
\\  
\hline\hline
 & ~ ~$g=0$ ~ ~ & $g=2$  & $g=0$ & $g=1$ & $g=2$ \\ 
\hline & & & & & \\[-2.5mm]
$T$& $~~\displaystyle \frac{a_{\rm U}}{2\pi}~~$  & $~~\displaystyle \frac{a_{\rm U}}{2\pi}~~$ & $\displaystyle ~~ \frac{a_v}{\pi}~~$ & $~~\displaystyle \frac{a_v}{2\pi}~~$ & $\displaystyle ~~ \frac{a_v}{\pi}~~$
\\[3mm] \hline
statistics & boson & fermion &  fermion  &  fermion  & boson 
\end{tabular}
\caption{\label{tab:accelT} Relation between an accelerated observer in quantum vacuum (Unruh case) to quantum vacuum accelerated by external field (M\"uller {\it et al}. case).}
\end{table}

\section{Observables}
We discuss   two observable effects inherent in $\Veff$: spontaneous pair production and light-by-light scattering.  Experiments seeking either of these effects  may one day help resolve the question of whether or not the two cases, accelerated observer and accelerated vacuum, lead to different physics.

The analyticity of quantum field theory demands that aside from heat fluctuations the accelerated observer also sees a rate of real $e\bar e$-pair production. Assuming that $g=1$ provides an accurate model of the physics seen by an accelerated observer,  pair production in this case   is obtained  according to Heisenberg-Euler-Schwinger for $g=1$ with the field strength written in terms of acceleration.  On the other hand, a strong field applied to the vacuum is expected to  produce the usual $g=2$ pair production~\cite{Narozhnyi70,Mueller75,Casher79,BialynickiBirula:1991tx,Kim:2003qp,Cohen:2008wz,Mihaila:2009ge,Dunne:2010zz,Labun:2010wf}.

We obtain the rate (per unit volume) of spontaneous field decay by pair emission, an effect possible only in the presence of an electric field, equivalently whenever the field invariant $a>0$, see \req{invar}.   The decay rate is controlled by the imaginary part of $\Veff$, which arises from the poles in the integrand of \req{Veff} at $u=in\pi/e\calE$ for integer $n$ [or equivalently in \req{Veff-step1} at $u=1$].  The integration contour is defined as in \req{m2minus} by assigning a small imaginary constant to the mass before rotating onto the positive real $u$ axis.  For the electric-only field
\beqn\label{Imagggen}
\Imag \Veff = \frac{\gamma_sm^2\TEH^2}{32\pi}\sum_{n=1}^{\infty}\frac{(-1)^n}{n^2}\cos(n\pi\fracg)\,e^{-n m/\TEH},\ |g|\le 2.
\eeqn
The total probability per unit volume per unit time of decay of the field is twice this imaginary part, $d\Gamma/d^4x=2\Imag\Veff$.

\begin{figure}
\includegraphics[width=0.48\textwidth]{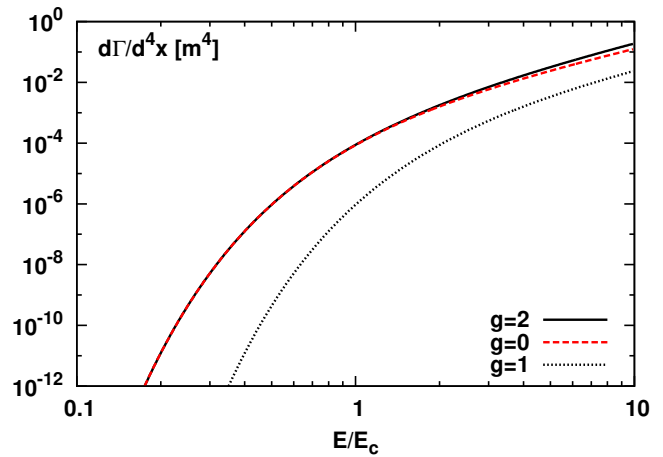}
\caption{The rate per unit volume of decay of the field $d\Gamma/d^4x=2\Imag\Veff$ with $\Imag\Veff$ given by \req{Imagggen}.  The electric field magnitude is normalized to $\calE_c=m^2/e$ the critical field strength, at which $\TEH\to m/\pi$.  For $g\neq 2$ the rate of field decay is reduced with the largest reduction for $g=1$.  Above $\calE_c$ we see suppression due to the $g$ factor modifying weights in the sum in \req{Imagggen}. \label{fig:ImV}}
\end{figure}

Setting $g=1$ (accelerated observer case) changes the analytic structure of $\Veff$, giving odd-$n$ terms in the sum zero weight.  The argument of the exponential is thus doubled,
\beqn\label{Imagg1}
\frac{d\Gamma}{d^4x}=\left.\!2\Imag \Veff\right|_{g=1}\!\! = \frac{\gamma_sm^2\THU^2}{16\pi}\sum_{n=1}^{\infty}\frac{(-1)^n}{n^2}e^{-n m/\THU}.
\eeqn
This change is especially visible in the rate per unit volume of particle emission $d\langle N\rangle/d^4x$, which is given by the first term of the series in \req{Imagggen}~\cite{Cohen:2008wz}.  
Because the $n=1$ term vanishes in Eq. (14), the $n=2$ term becomes the first term in the series, and $2/T_M=1/T_U$ appears in the exponent, corresponding to half the temperature value,
\beqn\label{dNd4xg1}
\left.\frac{d\langle N\rangle}{d^4x}\right|_{g=1}\!\! =\frac{\gamma_sm^2\THU^2}{32\pi}e^{-m/\THU}.
\eeqn
This notably shows the same numerical factors as the analogous result for $g=0,2$  after substitution of the Unruh temperature $\THU=\TEH/2$, as can be expected considering the analytic properties of the effective action \req{Veffggen} and the $g=1$ form \req{Veffg1}.

Figure \ref{fig:ImV} shows \req{Imagggen} for the  values  $g=0;1;2$. The results for $g=0;2$ are very similar and yield the largest total decay probability as a function of $g$. The reduction in the rate driven by the effective temperature parameter is largest for the particular case $g=1$.  Due to the exponential dependence, the reduction in the temperature parameter by a factor of 2 reduces spontaneous pair production below the critical field $\calE_c=m^2/e$ by many orders of magnitude. 

The real part of $\Veff$ leads to the  nonlinear field-field interaction.  For $g=1$ one finds
\beqn\label{LbLg1}
\Veff\Big|_{g=1}\!\!\simeq \frac{\gamma_s}{32\pi^2}\frac{e^4}{m^4}\frac{-1}{5760}\left(7(\vec B^2-\vec E^2)^2+4(\vec E\cdot\vec B)^2\right)
\eeqn
Terms containing higher powers of the field invariants are given in~\cite{Rafelski:2012ui}.  Relative to the $g=2$ values, the coefficients of $(\vec B^2-\vec E^2)^2$ and $(\vec E\cdot\vec B)^2$ in \req{LbLg1} are opposite in sign and suppressed: for light-by-light scattering experiments the important $(\vec E\cdot\vec B)^2$ term is 224 times smaller. 
\\[-0.2cm]

\section{Discussion and conclusions}
In a constant electric field ${\calE}\equiv a>0$, the electron fluctuations display a thermal Bose spectrum with temperature $\TEH=e\calE/m\pi=a_v/\pi$.  This result contrasts with the Fermi spectrum and the Unruh temperature $\THU=a_{\rm U}/2\pi$ experienced by an accelerated observer.  We discovered and exploited the  coincidence that case $g=1$ used in an accelerated vacuum produces physics relevant to the case of an  accelerated observer.
It is important to recognize that we have not, and in general cannot resolve the question of why we should or should not expect that the two cases, accelerated observer and accelerated vacuum,  to yield different or the same physics. 

We have   evaluated  the effective QED potential of a $g=1$ ``electron'' in presence of a constant electric field~\req{Veffg1}  finding   the form of the  QED effective  potential with the Unruh temperature and fermionic statistics appropriate for the physics of an observer accelerated in the electromagnetic force field.   Considering the quantum fluctuations of a ``classical spinning particle'' $g=1$  thus describes the Unruh result within the effective Heisenberg-Euler-Schwinger action. We argued that the  computation with $g=1$  is providing the complete effective potential generating the physics of an accelerated observer.

Two effects could be used to distinguish the accelerated observer with $g=1$ from the QED vacuum at $g=2$: $e\bar e$-pair  production in strong electric fields and nonlinear field-field interaction.  We have shown that both are greatly suppressed in the case $g=1$ relative to the QED $g=2$ expectation. QED strong field experiments such as light-field scattering~\cite{Rikken:2001zz,QAexperiment} will, if the accelerated observer case prevails, be seeking a much weaker signal. 

This proves the measurability of the difference between the frames down to arbitrarily small acceleration.  Being able to determine which is accelerated means that there is a universal class of inertial reference frames.  Introduction of a class of inertial reference frames realizes Einstein's interpretation of Mach's principle within the quantum theory.  The Einstein-Mach principle is incorporated in both the Unruh-type calculation (by comparing to the vacuum of flat Minkowski space) and the QED effective action (by renormalizing with respect to the zero-particle no-field state).

No experiment has yet tested macroscopic properties of the QED vacuum associated with the critical field strength $\calE_c=m^2/e$, a value considerably beyond the limiting field of Born-Infeld theory~\cite{Born:1934gh} and even beyond limits set considering precision strong field tests~\cite{Rafelski:1973fm}. For this reason it is necessary to ascertain that QED of strong fields, which differs from the expectations based on equivalent accelerated observer, is indeed different.  

Should  the strong-field QED experiment observe the original $g=2$ results, one would infer  a difference in temperatures \req{TUH} and \req{TEH}, and it follows that the two views of acceleration are not equivalent for any magnitude of the acceleration.  Note that the limit of weak acceleration is achieved in QED by considering fields smoothly varying on compact spatial domain.  On the other hand, the authors are not aware of a treatment of the Unruh detector in which the accelerated observer is smoothly connected to asymptotic inertial frames.  If one insists on the equivalence of the accelerated observer and the accelerated vacuum, our result, therefore, suggests that there is additional, undiscovered physics content in the properties of the Unruh accelerated detector.  

\section*{Acknowledgments} We thank B. M\"uller for his interest. This work was supported by a grant from the  U.S. Department of Energy No. DE-FG02-04ER41318.

\end{document}